\documentclass{elsart}
\usepackage{amssymb}
\usepackage{graphicx}
\usepackage{dcolumn}
\usepackage{bm}

\begin{document}
\begin{frontmatter}

\title{Neural Networks as a Composition Diagnostic for Ultra-high Energy Cosmic Rays}
\author{Andr\'e K. O. Tiba\corauthref{cor1}}
\author{, Gustavo A. Medina-Tanco}
\corauth[cor1]{tiba@astro.iag.usp.br}
\address{
Instituto de Astron\^omia, Geof\'isica e Ci\^encias Atmosf\'ericas \\
Universidade de S\~ao Paulo\\
Caixa Postal 9638 - CEP 05508-900, S\~ao Paulo - SP, Brasil }
\author{Sergio~J.~Sciutto}
\address{
Departamento de F\'isica and IFLP/CONICET, Universidad Nacional de
La Plata, \\ C. C. 67, 1900 La Plata, Argentina}

\begin{abstract}
We analyze here the possibility of studying mass composition in
the Auger data sample using neural networks as a diagnostic tool.
Extensive air showers were simulated using the AIRES code, for the
two hadronic interaction models in current use: QGSJet and Sibyll.
Both, photon and hadron primaries were simulated and used to
generate events. The output parameters from the ground array were
simulated for the typical instrumental and environmental
conditions at the Malarg\"ue Auger site using the code SAMPLE.
Besides photons, hydrogen, helium, carbon, oxygen, magnesium,
silicon, calcium and iron nuclei were also simulated. We show that
Principal Components Analysis alone is enough to separate
individual photon from hadron events, but the same technique
cannot be applied to the classification of hadronic events. The
latter requires the use of a more robust diagnostic. We show that
neural networks are potentially powerful enough to discriminate
proton from iron events almost on an event-by-event basis.
However, in the case of a more realistic multi-component mixture
of primary nuclei, only a statistical estimate of the average mass
can be reliably obtained. Although hybrid events are not
explicitly simulated, we show that, whenever hybrid information in
the form of $X_{max}$ is introduced in the training procedure of
the neural networks, a considerable improvement can be achieved in
mass discrimination analysis.
\end{abstract}

\end{frontmatter}

\section{\label{introduction}Introduction}

\par The origin of ultra-high energy cosmic rays (UHECRs) is currently
one of the most exciting problems in modern astrophysics. Up to
now, no astrophysical object is known that could accelerate
charged particles to such energies. Besides, if the sources are
located on cosmological distances, then it would be expected that
the Cosmic Rays arriving to the Earth would loose energy after
interacting with the cosmic microwave background above an energy
of about $6\times 10^{19}$ eV \cite{GZKcutoff}, the threshold for
photo-pion production in interactions with the cosmic microwave
background radiation (CMBR). This energy would therefore mark a
sharp bend in the Cosmic Rays energy spectrum. Is is not clear at
present whether such bent is seen by the experiments operating so
far. If the sources are nearby, then an anisotropic distribution
of arrival directions is expected because in this case the
directions of arrival would point to the sources, and this does
not seem to be the case either.

Alternative explanations of the existence of the UHECRs have been
developed over the last few years. Indeed, a whole spectrum of
models exists, ranging from the more conservative bottom-up
models, in which nuclei are accelerated up to the highest observed
energies \cite{Hillas84}, to the more radical top-down scenarios
where either new particles or exotic phenomena are invoked to
produce the particles at already higher energies than those
measured (\cite{Berezinsky97} for example).

The discrimination among these models, requires an accurate
determination of the energy spectrum, the distribution of arrival
directions and the identity of the particles.

Additionally, accuracy must be complemented with high statistics.
During the next few years, it is expected that at least two major
experiments will attain this requirements: Auger South and North
\cite{auger} and the Extreme Universe Space Observatory, EUSO
\cite{EUSO}.

In this work will concentrate on the specific characteristics of
the Auger experiment. Nevertheless, the following analysis might
be conceptually extended to any other air-shower-based ultra-high
energy cosmic ray experiment.

The Auger experiment consists of two detectors of about $3000$
km$^2$ each, located at sites in the Southern and Northern
hemispheres respectively. The Southern observatory is being
deployed at present in Malarg\"ue, Argentina, and is already
taking data as construction proceeds.  Each detector will be
capable of measuring the properties of the showers generated by
the ultra high energy cosmic rays. An array of surface detectors
(SD) will measure the lateral distribution of the shower at ground
level, while fluorescence detectors will measure the longitudinal
distribution of the shower traversing the atmosphere.

The development of extensive air showers (EAS), as characterized
by their lateral distribution, curvature of the shock front,
rising time, pulse shape, total number of photoelectrons, etc.,
carries information regarding the direction, energy and identity
of the incoming primary. However, while direction and energy can
be estimated rather easily from ground array data (e.g.
\cite{Billoir2000}), the definition of a convenient and efficient
diagnostic for primary identity discrimination remains a
challenging issue.

In particular, besides some punctual indications against UHE
photons as primaries \cite{Bird95,Halzen95,Nagano99}, only one
comprehensive study limiting the photon flux above $10^{19}$ eV
has been published \cite{Ave00} up to now, and it is based on
measuring the ratio of vertical and inclined showers at Haverah
Park (zenith angles $>60^{o}$). The separation between light
(protons) and heavier (Fe nuclei) hadrons is still much more
difficult.

Given the present uncertainties, results so far remain mostly
qualitative and it is likely that such a complex problem will not
be solved by the use of a single technique.

In this paper we present the results of an ongoing effort to
develop primary identification diagnostics with the aid of
multivariate and neural network techniques. A pragmatic approach
is taken to the practical problem of statistically determining the
identity of the primaries starting EAS at the top of the
atmosphere with the ground array of the Auger Observatory as the
specific target.

The paper is structured as follows. Section two describes the data
set, built from simulations of extensive air showers and their
surface detection. In section three we show and apply the
Principal Component Analysis (PCA) method to separate photons and
hadrons. In section four after a summary introduction to Neural
Networks (NNs) we apply them to the classification of proton and
iron nuclei (on an event-by-event basis) and to the problem of
classification in the case of a continuous mass spectrum which is
approximated by an eight nuclei cosmic ray flux. Section five
summarizes our results and conclusions.

\section{\label{secII} EAS simulations and detection}

A large sample of showers for primary photons and hadrons was
generated with the AIRES code and, transformed into ground array
events of a model Auger Observatory, simulated with the SAMPLE
code \cite{Billoir2000}.

The AIRES system is one of the widely used codes for EAS
simulation currently in use. All the relevant particles and
interactions are taken into account during the simulations, and a
number of observables are measured and recorded, among them, the
longitudinal and lateral profiles of the showers, the arrival time
distributions, and detailed lists of particles reaching ground
that can be further processed by detector simulation programs. The
AIRES system is explained in detail elsewhere \cite{aires0}.

The showers processed in this work consist of: (i) a series of
1831 proton, gamma, and iron showers, processed with AIRES 2.4.0
and QGSJet 98 hadronic interaction model, with energies in the
range $10^{17.5}$ eV to $10^{20.5}$ eV, and zenith angles in the
range 0 to 60 degrees; (ii) a series of 10,000 proton, He, C, O,
Mg, Si, Ca and Fe showers, with energies in the range $10^{17.5}$
eV to $10^{20.9}$ eV and zenith angles in the range 0 to 84
degrees, processed with AIRES 2.5 and both QGSJet 01 \cite{QGSJET}
and Sibyll 2.1 \cite{SIBYLL}. Each shower was reused 20 times at
different location in the array, and so the final number of
available events is $36,620$ for set (i), and $200,000$ for set
(ii). The surface detectors have been simulated using the SAMPLE
SD simulation program.

The directly observable outputs for each event, which include the
number and spatial distribution of triggered tanks and the time
profile of the signal at each station (the fluency times $T_{10}$,
$T_{50}$ and $T_{90}$), together with elementary reconstructed
quantities (e.g., primary energy $E_{\circ}$, zenith angle
$\theta$) are used to define different sets of parameters. The
energy and zenith angle were not reconstructed, but the input
parameters to shower simulations were directly used.

The number of ground events (measured with only the surface
detector array) are roughly ten times the number of hybrid events
(measured with both techniques, surface and florescence).
Therefore, it is important to evaluate the improvement on
classification efficiency obtained from the use of hybrid events.
A good idea on that respect can be obtained, without resorting to
full simulation of the longitudinal development reconstruction, by
simply adding the $X_{max}$ parameter to the set of ground array
parameters. Hence, whenever we want to assess the potential of
hybrid events, we include the $X_{max}$ value calculated by AIRES
without performing the fluorescence reconstruction of the showers.

\section{\label{secIII}Principal Component Analysis: photon-hadron separation}

The full set of directly measured and reconstructed quantities,
can be combined to form an \emph{n}-dimensional orthogonal
parameter space. This space can be further studied by using
Principal Component Analysis (PCA), in search for primary
separation.

The PCA method (for example, \cite{Hassoun95}) simply performs a
rotation in the \emph{n}-dimensional space to a new orthogonal
coordinate system whose unit vectors are the eigenvectors of the
system. These new axis have a special meaning, since their
associated eigenvalues are a measure of the dispersion of the data
along each axis. Thus, the principal eigenvector has the largest
associated eigenvalue, and therefore the largest dispersion, or
information content, of the sample; the second eigenvector has the
second largest dispersion and so on. Typically, one can quantify
the amount of information associated with a subset of axis, and
can even expect to uncover the true dimensionality of the system
if this has been overestimated.

One advantage of the PCA method is that, involving only rotations,
the new axis are only linear combinations of the original
magnitudes.

As an illustrative example \cite{GMTSci2001}, lets take a
parameter space defined arbitrarily by:

A (sort of) curvature estimator,

\begin{equation}
P_{1}= \left[ \frac{\langle T_{0,ext} \rangle - \langle T_{0,int}
\rangle}{\langle r_{ext} \rangle - \langle r_{int} \rangle}
\right] \times \sin{\theta} \label{eq01}
\end{equation}

\noindent where the subscripts "ext" and "int" refer to stations
that are farther away and nearer the shower axis than the median
distance $r_{c}$ of the triggered stations, and $r_{ext}$ and
$r_{int}$ are the average distances inside each region.

The third largest total number of vertical equivalent muons,

\begin{equation}
P_{2}= \left[ (vem)_{total} \right]_{3rd}\label{eq02}
\end{equation}

The pulse shape/rising time (average),

\begin{equation}
P_{3} = \langle \frac{T_{50}}{T_{10}+T_{50}} \rangle \label{eq03}
\end{equation}

\noindent where $T_{i}$ are the fluency times for 10\% and 50\% of
the total fluency at a given station.

The pulse shape/rising time (3rd largest value),

\begin{equation}
P_{4} = \left( T_{10}+T_{50}+T_{90} \right)_{3rd}\label{04}
\end{equation}

A lateral distribution estimator,

\begin{equation}
P_{5} = \left[ \frac{(vem)_{total}}{P_{4}} \right]_{5th} /
        \left[ \frac{(vem)_{total}}{P_{4}} \right]_{3rd}
        \label{eq05}
\end{equation}

The 3rd largest value of the rising time,

\begin{equation}
P_{6}=\left( \frac{T_{10}-T_{0}}{T_{90}-T_{0}}  \right)_{3rd}
\label{eq06}
\end{equation}

\noindent plus the median of the station distances to the axis of
the shower $P_{7}=r_{c}$, the primary energy $P_{8}=E_{0}$, the
zenith angle $P_{9}=\theta$ and the number of triggered stations
$P_{10}=N_{stat}$.

All these parameters are normalized later so that their dynamical
ranges are in the interval $(-1,1)$.

When a PCA analysis is performed in this parameter space for the
simulation set (i) of proton, iron and gamma primaries, it is
found that the first 4 eigenvectors are responsible for $\sim
80$\% of the variance (or information content) of the system. The
7th eigenvector is responsible for only $\sim 6$ \% of the
variance.

In this particular parameter space, the best separation between
nuclei and photons is obtained for the projection onto the plane
defined by the first and seventh eigenvectors (see Fig.~1). The
thick line, $EV_{7} = -48.89 \times (EV_{1}+0.007)^{2}+0.011$, in
Fig.~1 is able to separate very well nuclei from gamma
populations: the probabilities of misidentification are $\sim 3$\%
and $\sim 4$\% for photons and nuclei, respectively.

\begin{figure}[ht!]
\centering
\includegraphics[width=8.5cm]{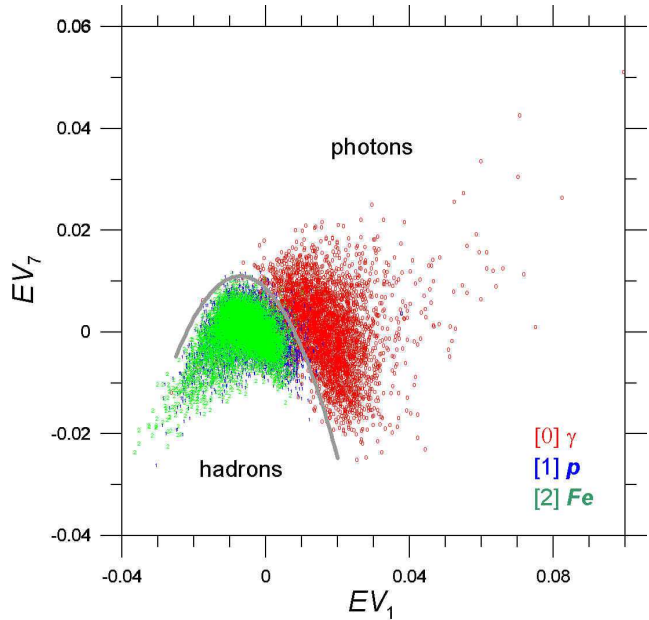}
\caption{PCA results on the illustrative parameter space. The best
separation between nuclei and photons is obtained for the
projection on the plane defined by the first and seventh
eigenvectors. The thick line misclassifies $4$\% of the nuclei as
photons and $3$\% of the photons as nuclei.}
\end{figure}\label{PCA01}

\begin{figure}[ht!]
\centering
\includegraphics[width=8.5cm]{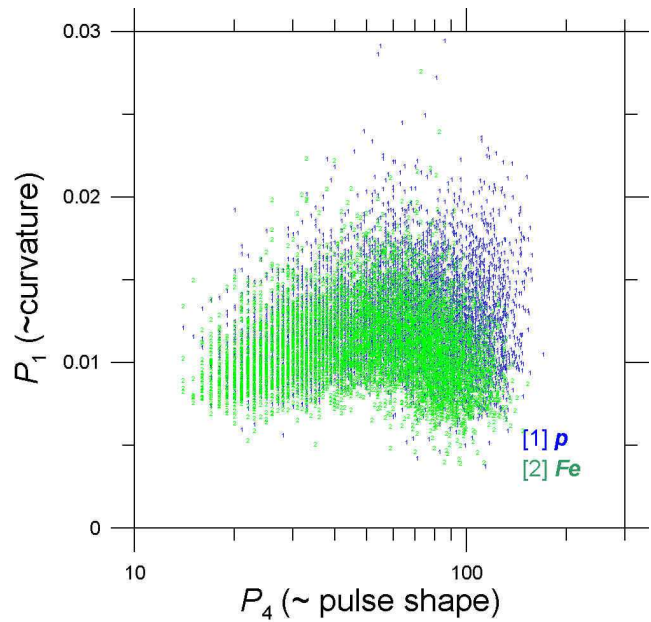}
\caption{Projection onto the $P_{1}$--$P_{4}$ plane of the sample
points, once the photon events have been extracted, showing the
difficulty involved in the separation of light and heavy nuclei.}
\label{PCA02}
\end{figure}

Once the photons have been separated, the same process could be
applied, in principle, to the further separation of nuclei among
themselves. However, this is a much more complicated problem as
can be seen in Fig.~1 or even in Fig.~\ref{PCA02}, which show that
proton and iron nuclei are truly mixed in these parameter space
regardless of what projection is chosen for the analysis. The
necessity of a more powerful technique is obvious and, as we will
show in the following section, a potentially useful diagnostic
tool can be obtained by resorting to neural network simulations.

\section{\label{secIV}Neural Networks applied to nuclei separation}

An alternative approach for hadronic primary separation can be
obtained by applying neural network technics to the problem.

\subsection{Neural Network design}

A neural network (NN), is structured in parallel layers of
neurons, connected to neurons in adjacent layers. Each connection
has a statistic weight (or just weight) since two neurons of the
same layer, should not ``see" the same input coming from a neuron
of the previous layer.

\begin{figure}[ht!]
\centering
\includegraphics[width=8cm]{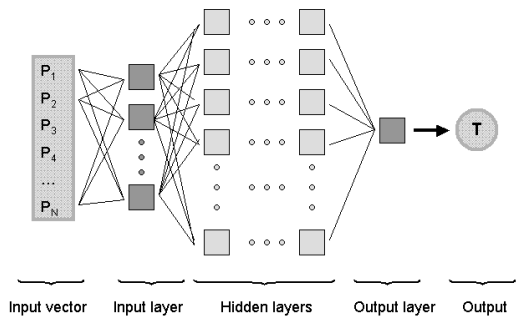}
\caption{A general network, with an input layer, an hidden layer
(multi-layer) and an output layer. Each square box represents a
neuron unit showed in Fig.~\ref{singleneuron}.}
\label{MultiLayerneuron}
\end{figure}

In general terms, a NN is formed by: the input layer, which is
connected to the input data vector; an indefinite number of hidden
layers and the output layer, the last layer of neurons (see
Fig.~\ref{MultiLayerneuron}).

\begin{figure}[ht!]
\centering
\includegraphics[width=7cm]{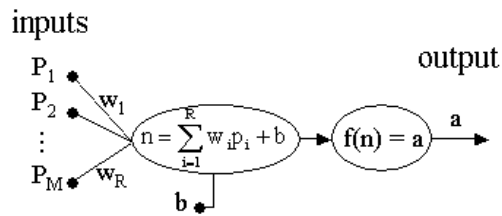}
\caption{A single neuron from Fig.~\ref{MultiLayerneuron}: the
input vector {\bf p}, the weights $w_{i}$ $(i=1,...,R)$, the bias
\emph{b}, the transfer function \emph{f} and the output signal
\emph{a}. The weight $w_{i}$ shows which importance each input
signal has for the neuron.} \label{singleneuron}
\end{figure}

The elementary processing unit in a NN is the artificial neuron.
The information arrives to the neuron from many input channels.
The information coming from each channel is linearly transformed
by applying a multiplicative weight and an additive bias and fed
to a transfer function which gives the neuron output signal (see
Fig.~\ref{singleneuron}). The bias parameter \emph{b} regulates
the threshold activation of the neuron (for boolean neurons). But
in our case, the bias is one more parameter to be adjusted by the
training algorithm.

The training algorithms that we tested, required neurons with
differentiable transfer functions. After tests with several
transfer functions, we selected basically two types: (a) linear:

\begin{equation}
f(n) = n \label{eq_linear}
\end{equation}

\noindent and (b) hyperbolic tangent sigmoid:

\begin{equation}
f(n) = \left [ \frac{e^{n} - e^{-n}}{e^{n} + e^{-n}}\right]
\label{eq_tangente}
\end{equation}

\noindent where:

\begin{equation}
n = \sum_{i=1}^{R}w_{i}p_{i} + b
\end{equation}

In our case, the last layer of all networks has one neuron with
linear transfer function. All other neurons of the network have
hyperbolic tangent sigmoid as transfer function.

For simplicity, we adopt hereafter the following notation in order
to represent a network. The notation 7i-10-10-10-1o, for example,
means: one input data vector of dimension 7; three layers, each
one with 10 neurons (all hyperbolic tangent sigmoid), and the
output layer with one neuron(linear). This network has 311 free
parameters: 280 weights and 31 biases. Networks of very different
architectures and sizes were tested, running from approximately 80
to more than 1100 weights.

The adopted solution was a supervised training algorithm. The
training data set $\Phi$, is a collection of \emph{Q} events,
$\Phi = \{\{\textbf{p}_{q},y_{q}\}\}_{q=1}^{Q}$ where
$\textbf{p}_{q}$ is the input data vector (the measured parameters
that characterize the event) and $y_{q}$ is the identity of the
event (atomic mass of the primary nucleus). The supervised
training minimizes the difference between the desired output
$y_{q}$ and the computed output $a_{q}$, adjusting the weights and
biases of the NN.

The minimization function is generally the square error function

\begin{equation}
F(\textbf{x}) = \sum_{q=1}^{Q}(y_{q}(\textbf{x}) - a_{q})^{2}
\label{eq_mse}
\end{equation}

\noindent where $\textbf{x}$ is the vector formed by the weights
and biases of the NN.

The basic \emph{backpropagation} algorithm \cite{rumc86} was
developed for networks with multiple layers of neurons. This is a
feedforward algorithm, whose main characteristic is that the
process of weight adjusting progresses from the last to the first
neuron layer. Some of the backpropagation algorithms we tested use
equation (\ref{eq_mse}) as their minimizing function.

A frequent problem in network training is the lost of
generalization capability. The algorithm attains a good training
level (i.e., low values of $F(\textbf{x})$) but it is not capable
of reproducing the same good performance when faced to an
independent test set. This problem can often be traced to the
development of some few large weights during the process of
minimization of the error function. These large weights are
responsible, in turn, for a very sensitive NN, which can react
unpredictably when presented with inputs that depart, even
slightly, from the original training set.

An attempt to solve this problem, used in the present work, is the
Bayesian Regularization Backpropagation Algorithm (BRBA),
described by \cite{Demuth99} and detailed in \cite{foha97,hade96}.
Its minimization function is a modified version of (\ref{eq_mse}),
aimed at improving the network generalization capability:

\begin{eqnarray}
F(x(k))&&=\alpha(k)\sum_{q=1}^{Q}(y_{q} - a_{q}(k))^{2} +
\beta(k)\sum_{i=1}^{\Omega}x(k)^{2}\nonumber\\
&&=\alpha(k)E_{er}(k) +\beta(k)E_{x}(k)\label{funcao_GN}
\end{eqnarray}

\noindent where $\Omega$ is the number of weights of the NN,
$\alpha$ and $\beta$ are parameters that change at each iteration
$k$.

Roughly speaking, the term $E_{er}$ is responsible for the network
learning, while $E_{w}$ is correlated with its generalization
capability because it keeps checked the value of sum of the square
of the  weights. By minimizing both terms simultaneously, $E_{er}$
and $E_{w}$, one expects an equilibrium between learning and
generalization capability.

In all cases after training our networks with a given training
data set, their generalization capability was checked with an
equivalent independent control set.

\subsection{Two components UHECR flux: p-Fe separation}

Given a specific problem, there are no rules to design on optimal
NN. Each problem is a case study in itself. Therefore, many
different NN architectures were tested. Several backpropagation
algorithms, number of layers, number of neurons in each layer,
transfer functions, number of iterations and input sets of
parameters were tried until acceptable results were attained
\cite{GMTSciTiba2003}.

\begin{figure}[ht!]
\centering
\includegraphics[width=10.cm]{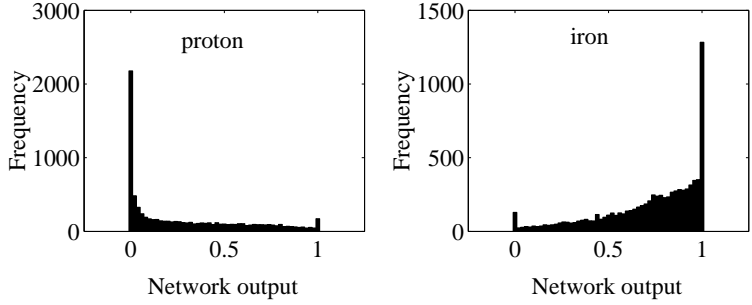}
\caption{Results of the application of a trained feedforward
network to an independent control sample of 13000 events triggered
by protons and iron nuclei. The network was trained to output a
value of zero (one) for a proton(iron) primary. Tails, therefore,
correspond to misclassified events. Only surface array information
was included.}\label{p-fe_nXmax}
\end{figure}

The best results were obtained by first transforming the parameter
space into its eigenvector space employing PCA techniques
(reducing at the same time the dimensionality of the input data).
As a first example, in Fig.~\ref{p-fe_nXmax}, we show the results
for a feedforward network 7i-10-10-10-1o.

The input parameters used, $P_{i}$, based in direct observables
and reconstructed magnitudes from the surface array detector:

\begin{eqnarray}
P_{1}= \left ( \frac{1}{N_{stat}} \right) \sum_{i=1}^{N_{stat}}
(vem)_{i} \left( \frac{r_{0,i}}{1000m} \right)^{3}\
\label{eq_p-Fe_P1}
\end{eqnarray}

\begin{eqnarray}
P_{2}= \left ( \frac{1}{N_{stat}} \right) \sum_{i=1}^{N_{stat}}
T_{sp,i} \left( \frac{r_{0,i}}{1000m} \right)^{-2}\,
\label{eq_p-Fe_P2}
\end{eqnarray}

\begin{eqnarray}
P_{3}= \left ( \frac{1}{N_{stat}} \right) \sum_{i=1}^{N_{stat}}
T_{10,i}, \label{eq_p-Fe_P3}
\end{eqnarray}

\begin{equation}
P_{4}= \left ( \frac{1}{N_{stat}} \right) \sum_{i=1}^{N_{stat}}
T_{50,i}, \label{eq_p-Fe_P4}
\end{equation}

\begin{equation}
P_{5}= \left ( \frac{1}{N_{stat}} \right) \sum_{i=1}^{N_{stat}}
T_{90,i}, \label{eq_p-Fe_P5}
\end{equation}

\noindent plus initial energy ($P_{6} = E_{0}$), zenith angle
($P_{7} = \theta$) and number of triggered stations ($P_{8} =
N_{stat}$); where, for the station \emph{i}: $(vem)_{i}$ is the
signal measured in vertical equivalent muons, $T_{sp,i}$ is the
arrival time of the shower plane, $r_{0,i}$ is the distance to the
shower axis and $T_{j,i}$ (j=10, 50, 90) are the rising times, at
which \emph{j}\% of the integral value of the signal is attained.

A set of cuts was applied to remove events too far from the
average behavior of the parameter distributions. The cuts,
described in Table~\ref{cutoof_1},  eliminate roughly $1/4$ of the
events.


\begin{table}[ht!]
\begin{center}
\renewcommand{\arraystretch}{1.5}
 \caption{The cuts eliminate roughly 1/4 of the events. The
remaining 3/4 events were randomly split into training and
independent control sets.} \label{cutoof_1} \vspace{0.1cm}
\begin{tabular}{c} \hline\hline
Cuts\\
\hline
\hbox{$40\leq P_{1} \leq 3\times10^{5}$} \\
\hbox{$7\times10^{-3}\leq P_{2} \leq 20$}
\\ \hbox{$3\times10^{-2}\leq P_{3} \leq 80$}
\\ \hbox{$3\leq$ P$_{5}\leq 250$} \\
\hbox{$P_{8} = N_{stat}\geq 4$} \\ \hline\hline
\end{tabular}\\
\end{center}
\end{table}

The network was trained to output desired $0$ for protons and $1$
for iron nuclei. The training set had 10000 events (5000 for each
kind of nucleus).

Fig.~\ref{p-fe_nXmax} shows the result of applying the trained
network to independent control samples of 13000 events each one.

If we assume that an output $\leq  0.5$ identifies the event as
proton generated and an output$> 0.5$ signals to an iron nucleus,
then the misclassification probabilities for proton and iron
nuclei are, respectively, $P_{p} \sim 27$\% and  $P_{Fe} \sim
16$\%.

\begin{figure}[ht!]
\centering
\includegraphics[width=10cm]{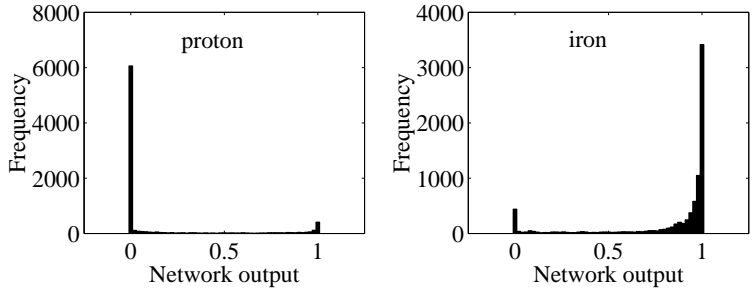}
\caption{Same as Fig.~\ref{p-fe_nXmax}, but now hybrid events were
considered (basically through the inclusion of $X_{max}$). A much
clearer separation is obtained, despite some events are still
misclassified.}\label{p-fe_Xmax}
\end{figure}

In order to asses the impact of using information coming from
hybrid events, we performed an additional run including also
$X_{max}$. The corresponding output is showed in
Fig.~\ref{p-fe_Xmax}. A noticeable improvement shows up clearly as
the misclassification probability goes down to $\sim 12$\% for
protons and $\sim 12$\% for iron. Furthermore, the number of
ambiguous events with intermediate results between 0 and 1
diminishes noticeably producing a cleaner output.

Since each shower was reused 20 times to simulate ground events,
those events may not be completely independents themselves. This
could result in some artificial improvement of the network
response. To check this possibility, we eliminated the shower
repetitions, lowering the number of available events by a factor
of 20, but guaranteing the independence of the events.

We use the same network architecture as before, but this time only
one cut was applied, $P_{8} = N_{stat} \geq 4$, since additional
cuts would eliminate too many events hindering the analysis. The
training set had 900 events (450 of each kind of nucleus) and the
independent control samples had roughly 480 events.

The misclassification probabilities for proton and iron nuclei
were, respectively, $P_{p} \sim 26$\% and $P_{Fe} \sim 20$\% for
surface events. The same misclassification probabilities, for
hybrid events, were  $P_{p} \sim 13$\% and $P_{Fe} \sim 13$\% for
proton and iron nuclei. Therefore, there was no artificial
improvement in network results because of the not fully
independence of the events in the former sample. Note also that
this result holds even when the second, completely independent
sample, had less cuts in parameter space, which should worsen the
results.

\begin{figure}[ht!]
\centering
\includegraphics[width=10cm]{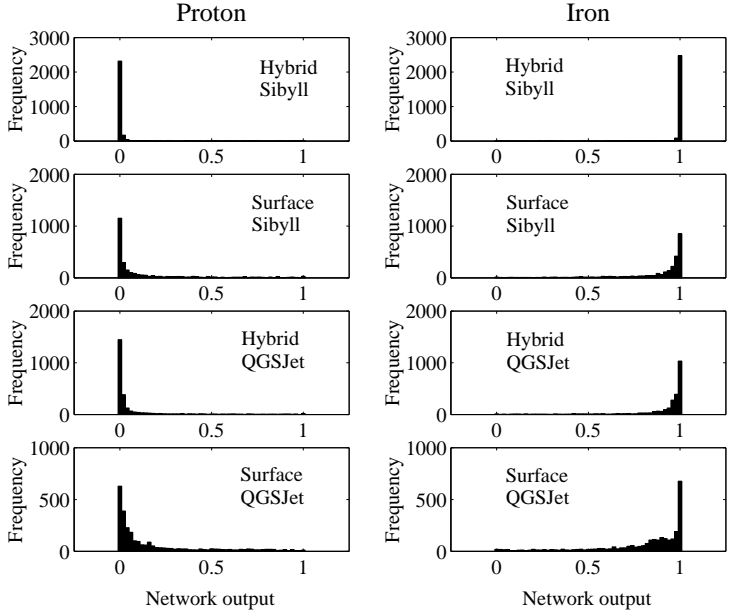}
\caption{The proton(iron) control samples simulated by the network
are in left(right) histograms. The top(down) four histograms were
simulated with Sibyll(QGSJet) events. The networks trained with
Sibyll events had thinner and clearer output distributions.}
\label{Fig.p-Fe_Hmodels_01}
\end{figure}

In order to maximize the number of events available, training and
control data sets of mixed events, of QGSJet and Sibyll
interaction models, were used in the previous examples, which
hampers the performance of the networks. From now on we show
networks simulated with either QGSJet or Sibyll events. Special
care was put in selecting equivalent samples of QGSJet and Sibyll
events with the same distribution in initial energy $E_{0}$ and
zenith angle.

The training set had 6000 events (3000 of each nucleus) and the
control independent samples had 2500 events of each nuclei. The
same network architecture 7i-10-10-10-1o and the same procedures
used before were applied. The results of simulating  QGSJet and
Sibyll events separately are shown in
Fig.~\ref{Fig.p-Fe_Hmodels_01}.

The four upper histograms are the outputs of networks simulated
with Sibyll control samples while the four lower histograms
correspond to networks simulated with QGSJet control samples.
Results are presented for surface and hybrid events. The networks
simulated with Sibyll events shows thinner pics around the desired
outputs (0 for proton and 1 for iron) and their distributions have
a smaller number of ambiguous events than in the case of QGSJet.

The misclassification probabilities of proton and iron nuclei,
with surface and hybrid events are in
Table~\ref{tableCP1_QG_Sb_selected}.


\begin{table}[ht!]
\begin{center}
\renewcommand{\arraystretch}{1.5}
 \caption{The misclassification probabilities of proton($P_{p}$)
and iron ($P_{Fe}$) nuclei for networks simulated with control
independent samples of different hadronic models.}
\label{tableCP1_QG_Sb_selected} \vspace{0.1cm}
\begin{tabular}{llcc}\hline \hline
\hbox{Hadronic model}& \hbox{Detection}& \hbox{$P_{p}$(\%)} &
\hbox{$P_{Fe}$(\%)}\\
\hline \noalign{\smallskip}
\hbox{Sibyll}& \hbox{Hybrid}& 1 & 1\\
\hbox{Sibyll}& \hbox{Surface}& 12 & 10\\
\hbox{QGSJet}& \hbox{Hybrid}& 8& 9\\
\hbox{QGSJet}& \hbox{Surface}& 15 & 14\\ \hline\hline
\end{tabular}
\end{center}
\end{table}

As expected, the results for the misclassification probabilities
are much better than in the previous examples, where mixed
hadronic interaction models were used.

Table~\ref{tableCP1_QG_Sb_selected} shows that the improvement in
misclassification probabilities when going from surface to hybrid
events is, for Sibyll events, one order of magnitude. For QGSJet
the improving is roughly a factor of two.

In the case of pure surface events, QGSJet and Sibyll
misclassification probabilities are approximately equivalent, with
a slight advantage in favor of Sibyll. Nevertheless, in the case
of hybrid events, those probabilities are very different, being a
factor of 10 smaller for Sibyll than for QGSJet. The QGSJet
hadronic model has a higher multiplicity than Sibyll, which
produces a larger $X_{max}$ dispersion ($\sigma_{X_{max}}^{2}$)
and more fluctuations \cite{Almu04}. Apparently, the networks had
more difficult in training and recognition of these noisier data
sets.

In order to study the effect of our lack of knowledge about the
true hadronic interaction model operating in nature at these
energies, we trained networks with events simulated with one
hadronic model and tested them afterwards with events simulated
with the other hadronic interaction model. The same network
architecture was used as before. The corresponding
misclassification probabilities of control samples are shown in
Table~\ref{tableCP1_QG_Sb_selected2}.


\begin{table}[ht!]
\begin{center}
\renewcommand{\arraystretch}{1.5}
 \caption{ The same as in Table~\ref{tableCP1_QG_Sb_selected}, but
different hadronic interaction models were used for training and
testing.} \label{tableCP1_QG_Sb_selected2} \vspace{0.1cm}
\begin{tabular}{lllcc}\hline \hline
\multicolumn{2}{c}{\hbox{Hadronic model}} & & &\\
\hbox{Trained}& \hbox{Tested} &\hbox{Detection} & \hbox{P$_{p}$(\%)}& \hbox{P$_{Fe}$(\%)}\\
\hline \noalign{\smallskip}
\hbox{Sibyll}& \hbox{QGSJet} &\hbox{Hybrid}& 18& 15\\
\hbox{Sibyll}& \hbox{QGSJet} &\hbox{Surface}&  27& 21\\
\hbox{QGSJet}& \hbox{Sibyll} &\hbox{Hybrid}&  6& 10\\
\hbox{QGSJet}& \hbox{Sibyll} &\hbox{Surface}&  21& 17\\
\hline\hline
\end{tabular}
\end{center}
\end{table}

Table~\ref{tableCP1_QG_Sb_selected2} shows the important result,
that NN should be able to classify primary identity correctly, at
least in 70 \% of the events, independently of the hadronic
interaction model used in the simulations.

Furthermore, if the true hadronic interaction model is unknown,
the safest bet is to train with QGSJet and, when possible, to
limit the analysis to hybrid events.

The tests performed on binary mixture of proton and iron nuclei
show that NN are capable of acceptable classification almost on an
event by event bases. However, a binary model for the flux of
UHECR, formed only by proton and iron nuclei, is not realistic. If
heavy nuclei, say iron, arrive at the top of atmosphere, lighter
nuclei should arrive too, even if only due to photodisintegration
via interactions with the infrared background in the intergalactic
medium. Therefore, we expanded the same NN analysis to a group of
eight different nuclei.

\subsection{Multi-component UHECR flux: 8 different nuclei}

A flat mass spectrum, composed by a mixture of 8 nuclei, was used
for training: $^{1}$H, $^{4}$He, $^{12}$C, $^{16}$O, $^{24}$Mg,
$^{28}$Si, $^{40}$Ca and $^{56}$Fe.

As in the previous case, we tested different NN architectures,
learning and training algorithms and sets of parameters. This
time, unfortunately, we were unable to design a NN capable of
adequately separating individual events.

Consequently, we limited our attempts to a statistical
characterization of the mass spectrum.

We were forced to apply additional cuts to the sample of events,
based on a series of curve fittings to the lateral distribution of
the individual showers. The curve fittings characterize the shape
of the lateral distribution function, the curvature of the shower,
and the radial dependence of the time structure of the shower
front:

\begin{equation}
\it (vem) - (vem)_{90,min}\ = a_{pe} \left ( \frac{r}{1000}
\right)^{b^{pe}}\label{eq_SP2_01}
\end{equation}

\begin{equation}
\it t_{00}\ = a_{0} \left ( \frac{r}{1000} \right)^{b^{0}}
\label{eq_SP2_02}
\end{equation}

\begin{equation}
\it t_{50} - t_{50,min}\ = a_{50} \left ( \frac{r}{1000}
\right)^{b^{50}}\label{eq_SP2_03}
\end{equation}

\begin{equation}
\it t_{90} - t_{90,min}\ = a_{90} \left ( \frac{r}{1000}
\right)^{b^{90}}\label{eq_SP2_04}
\end{equation}

The cuts were applied on the bases of the value of some of the
fitted parameters in eq. (\ref{eq_SP2_01})-(\ref{eq_SP2_04}). The
new cuts are given explicitly in Table~\ref{cutoof_2}. Again, only
high flyers were eliminated. Restrictions were also applyed on the
linear regression of the various fittings, $\rho_{i}$. We added an
energy cut in energy, similar to Auger energy threshold.


\begin{table}[ht!]
\begin{center}
\renewcommand{\arraystretch}{1.5}
 \caption{The cuts of Table~\ref{cutoof_1} and Table~\ref{cutoof_2},
 applied together, eliminate roughly 2/3 of the
events.}\label{cutoof_2} \vspace{0.1cm}
\begin{tabular}{ccccc} \hline\hline
& & Cuts & & \\
 \hline
\hbox{$E_{0} > 10^{19}$~eV} && \hbox{$-5\leq b_{pe,i} \leq 0$} &&
\hbox{$0\leq b_{0,i} \leq 7$}\\
\hbox{$0\leq b_{50,i} \leq 3$} && \hbox{$0\leq b_{90,i}\leq 2.5$}
&& \hbox{$\rho_{i} \geq 0.85$} \\ \hline\hline
\end{tabular}\\
\end{center}
\end{table}

Since event-by-event classification proved too unreliable, we
limited ourselves to the statistical determination of the average
atomic mass of different composition spectra. As an example we
used, arbitrarily, simulated composition spectra of the form
$dN/dA \propto A^{\nu}$, with $\nu = -0.5, 0.0, 0.5, 1.0$ and 2.0.

For each value of $\nu$, $1000$ mass spectra of 300 events each,
were used to calculate the distribution function of the estimated
average mass.

The median output of the network for each nucleus was re-scaled to
its actual atomic mass by fitting an exponential function. The
criterion used to select among different neural networks was,
simultaneously, the goodness of the exponential fitting and the
dynamical range of the output between proton and iron.

\begin{figure}[ht!]
\centering \vspace*{2.0mm}
\includegraphics[width=10cm]{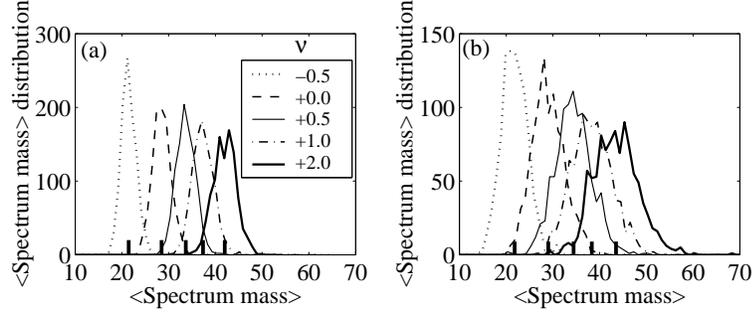}
\caption{Mass average for 1000 samples of 300 events each
simulated with network 7i-4-4-4-1o, for (a) hybrid events (i.e.,
with $X_{max}$) and (b) pure surface events (i.e., without
$X_{max}$). The vertical thick lines are, the true values of the
average spectral mass for each $\nu$.} \label{allnuclei}
\end{figure}

Fig.~\ref{allnuclei} shows the estimated average masses for
different spectral indices $\nu$ for both, hybrid (a) and pure
surface events (b). The vertical thick lines are the true values
of the average spectral mass for each $\nu$.

We used the set of parameters given by equations
(\ref{eq_p-Fe_P1}-\ref{eq_p-Fe_P5}) plus energy, zenith angle and
number of triggered stations. The network 7i-4-4-4-1o presented
the largest dynamical range between the proton and iron control
sample medians among all the network architectures tested. The
cuts of Table~\ref{cutoof_1} and Table~\ref{cutoof_2} reduced 2/3
the number of selected events. Therefore, we mixed our QGSJet and
the Sibyll samples first to keep a higher number of events for
training. This means that the results presented here are a worst
case scenario.

\begin{figure}[ht!]
\centering \vspace*{2.0mm}
\includegraphics[width=10cm]{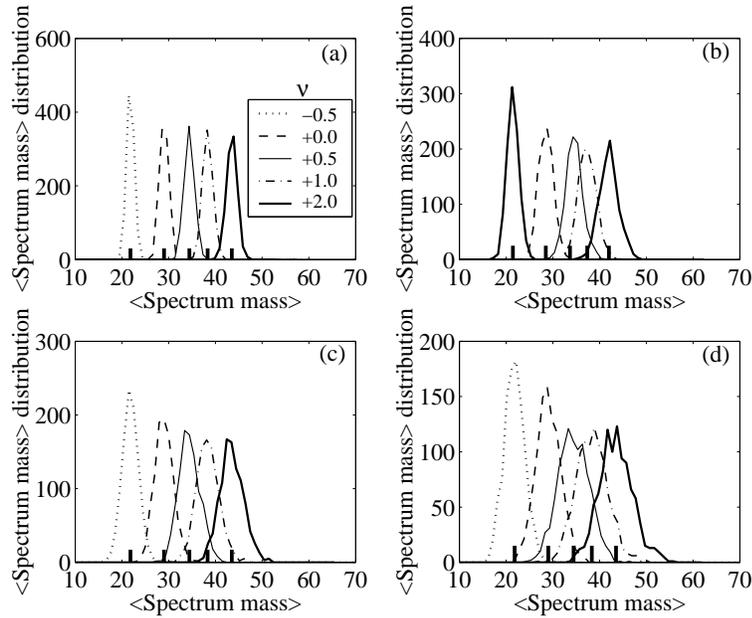}
\caption{The same of Fig.~\ref{allnuclei}. (a) - Sibyll, hybrid
detection; (b) - Sibyll, surface detection; (c) - QGSJet, hybrid
detection; (d) - QGSJet, surface detection. Statistical parameters
from these distributions are on Table~\ref{P6t_results}.}
\label{allnuclei_hmodel}
\end{figure}

The errors in the average value of the distributions is of the
order of 7\% for hybrid (Fig.~\ref{allnuclei}.a) and 12\% for
surface (Fig.~\ref{allnuclei}.b) events (see
Table~\ref{P6t_results}).


\begin{table}[ht!]
\begin{center}
\renewcommand{\arraystretch}{1.5}
\caption{Statistical parameters of distributions from
Fig.~\ref{allnuclei} and Fig.~\ref{allnuclei_hmodel}. $\langle
A_{t}\rangle$ is the true average distribution (vertical thick
lines), $\sqrt{\sigma^{2}}$ is the standard deviation error,
$\frac{\sqrt{\sigma^{2}}}{\langle A_{t}\rangle}$ is standart mean
error.} \label{P6t_results} \vspace{0.1cm}

\begin{tabular}{lcccccc}

& & & \multicolumn{2}{c}{\hbox{Hybrid detec.}} &
\multicolumn{2}{c}{\hbox{Surface detec.}}
\\\hline\hline

\hbox{H. model} & \hbox{$\nu$}& \hbox{$\langle A_{t}\rangle$} &
\hbox{$\sqrt{\sigma^{2}}$} & $\frac{\sqrt{\sigma^{2}}}{\langle
A_{t}\rangle}$\hspace{0.1cm}(\%) & \hbox{$\sqrt{\sigma^{2}}$} &
$\frac{\sqrt{\sigma^{2}}}{\langle A_{t}\rangle}$\hspace{0.1cm}(\%)\\
\hline \noalign{\smallskip}
\hbox{QGSJet}&-0.5&21.45& 1.59 &7 & 2.79 &13\\
\hbox{~~~+}&+0.0&28.50& 1.90&7 & 3.45 &12\\
\hbox{Sibyll}&+0.5&33.68& 2.06&6 & 3.96 &12\\
&+1.0&37.35& 2.28&6 & 4.29 &11\\
&+2.0&42.00& 2.42&6 & 4.82 &11\\
& &  & & & &\\
\hbox{Sibyll}&-0.5&21.45& 0.81 &4 &1.32 &6\\
&+0.0&28.50& 0.99&3 & 1.61 &6\\
&+0.5&33.68& 1.14&3 & 1.84 &5\\
&+1.0&37.35& 1.13&3 & 1.92 &5\\
&+2.0&42.00& 1.16&3 & 2.10 &5\\
& &  & & & &\\
\hbox{QGSJet}&-0.5&21.45& 1.74 &8 & 2.20 &10\\
&+0.0&28.50& 1.99 &7 & 2.64 &10\\
&+0.5&33.68& 2.28 &7 & 3.29 &10\\
&+1.0&37.35& 2.43 &7 & 3.40 &9\\
&+2.0&42.00& 2.55 &6 & 3.61 &9\\ \hline\hline
\end{tabular}
\end{center}
\end{table}

The same analysis, but discriminating by hadronic interaction
model is presented in Fig.~\ref{allnuclei_hmodel}. The training
set had 12,000 events (1,500 for each nucleus)  and control
samples had 2,000 events for each nucleus. The upper panels show
the networks trained and simulated with Sibyll events, while the
lower panels show the same but for QGSJet events.

As in the case of the proton-iron mixture, the networks were able
to classify better when trained with Sibyll events than with
QGSJet.

The errors on the determination of the average mass of the
distributions are given in Table~\ref{P6t_results}. For
Sibyll(QGSJet) events, they were of the order of 6\%(10\%), for
surface detection, and can go down by as much as a factor of two
for hybrid events.

With the aim of testing the effect of our lack of knowledge of the
true hadronic interaction model, we trained NN with events of one
interaction model and simulated them with events of the other
interaction model. The tested networks were the same as in the
previous example.

The networks trained with Sibyll events and tested with QGSJet
events presented smaller standard deviation and standard mean
error than those trained with QGSJet and simulated with Sibyll
(see, Table~\ref{P6t_results2}).

In the worst possible scenario, in which we have only surface
information and employ the wrong hadronic interaction model, our
neural networks can still calculate the average of the mass
spectrum with an error inferior to 20\%.

These results show, surprisingly, that NN are only weakly
dependent of the hadronic interaction model used.


\begin{table}[ht!]
\begin{center}
\renewcommand{\arraystretch}{1.5}
\caption{The same of Table~\ref{P6t_results}, but now the networks
were trained and tested with events of different hadronic models.}
\label{P6t_results2} \vspace{0.1cm}
\begin{tabular}{lrcccccc}

\multicolumn{2}{c}{\hbox{ Had. model}}&&&
\multicolumn{2}{c}{\hbox{ Hybrid det.}} &
\multicolumn{2}{c}{\hbox{ Surface det.}} \\ \hline\hline

\hbox{Trained}&\hbox{Tested}& \hbox{$\nu$}& \hbox{$\langle
A_{t}\rangle$} & \hbox{$\sqrt{\sigma^{2}}$} &
$\frac{\sqrt{\sigma^{2}}}{\langle A_{t}\rangle}$\hspace{0.1cm}(\%)
& \hbox{$\sqrt{\sigma^{2}}$} &
$\frac{\sqrt{\sigma^{2}}}{\langle A_{t}\rangle}$\hspace{0.1cm}(\%)\\
\hline \noalign{\smallskip}
\hbox{Sibyll}&\hbox{QGSJet}&-0.5&21.45& 2.43 &11 &3.58 &17\\
&&+0.0&28.50& 2.88&10 & 3.93 &14\\
&&+0.5&33.68& 3.02&9 & 4.21 &13\\
&&+1.0&37.35& 3.36&9 & 4.25 &11\\
&&+2.0&42.00& 3.38&8 & 4.38 &10\\
&& &  & & & &\\
\hbox{QGSJet}&\hbox{Sibyll}&-0.5&21.45& 3.43 &16 & 4.57 &21\\
&&+0.0&28.50& 4.98 &17 & 5.18 &18\\
&&+0.5&33.68& 5.94 &18 & 5.90 &18\\
&&+1.0&37.35& 6.43 &17 & 6.48 &17\\
&&+2.0&42.00& 6.91 &16 & 6.72 &16\\  \hline\hline
\end{tabular}
\end{center}
\end{table}

\section{\label{secV}Conclusion}

In the present work we showed the potential of both PCA and Neural
Networks applied to mass composition analysis of UHECRs.

In particular, we studied mass discrimination in three kinds of
cosmic ray mixtures: (a) photon-nuclei, (b) proton-iron and (c) a
mass spectrum comprising eight nuclei running from proton to iron.

In cases (a) and (b) we showed that event by event classification
was possible while, in a multi-component mixture, only a
statistical characterization of the mass spectrum is possible.

The PCA method was applied to the problem of separating photons
from hadrons. In this way, and for hybrid events, we were able to
attain a probability of misidentification of $\sim 3$\% for
photons $\sim 4$\% for hadrons (p and Fe). Nevertheless, we were
unable to separate nuclei among themselves by using PCA alone.
Neural networks were used to attack this problem.

In the case of binary mixtures of proton and iron, we were able to
design neural networks whose misclassification probabilities are
12\%(1\%) and 10\%(1\%) for proton and iron with surface(hybrid)
detection, for Sibyll training and testing. For QGSJet events, on
the other hand, the misclassification probabilities are 15\%(8\%)
and 14\%(9\%) for proton and iron with surface(hybrid) detection.
The QGSJet hadronic model generates noisier shower than Sibyll
because of its higher multiplicity. Apparently, the networks had
more difficulty in train and testing the noisier QGSJet event
sets.

Since the true hadronic interaction model at UHECR energies is
still very much under debate, we also trained networks with events
of one hadronic model and simulated them with an independent
control sample made up of events of the other hadronic model. The
tests showed that the networks are able to classify correctly, at
least 70\% of the events (iron or proton, surface or hybrid
detection), independently of the assumptions made on the hadronic
interaction model.

For a mixture of eight nuclei, we were unable to design a network
capable of discrimination on an event-by-event bases. Therefore,
we attempted only the statistical determination of the average
atomic mass of different mass spectra. As an example, we show
results here for composition spectra of the form $dA/dN \propto
A^{\nu}$, with $\nu = -0.5, 0.0, 0.5, 1.0$ and 2.0.

For networks trained with a mixture of QGSJet and Sibyll events,
the errors in the average of mass spectrum distributions were of
order of 12\% (7\%) for surface(hybrid) detection. When separating
the events by hadronic model, the errors in the determination of
the average mass of the spectra went down to $\sim$ 6\%(3\%) for
Sibyll events, and $\sim$ 10\%(7\%) for QGSJet events, for
surface(hybrid) detection.

A weak dependence with hadronic interaction model was confirmed
again for the NN, by training and testing the same network with
different hadronic interaction models. The errors incurred in the
determination of the average spectral mass, in this case, were
less than 20\%, for surface or hybrid detection regardless of the
assumed model.

\section{\label{secVI}Acknowledgments}
This work was supported by the Brazilian agencies CAPES, CNPq and
FAPESP. The authors thank Jeferson~A.~Ortiz and Vitor~de Souza for
useful discussions.

\bibliographystyle{elsart-num}

\begin{thebibliography}{10}
\expandafter\ifx\csname url\endcsname\relax
  \def\url#1{\texttt{#1}}\fi
\expandafter\ifx\csname
urlprefix\endcsname\relax\def\urlprefix{URL }\fi


\bibitem{GZKcutoff}
K. Greisen, Phys. Rev. Lett. 16 (1966) 748; G. T. Zatsenpin, V. A.
Kuz'min, Zh. Eksp. Teor. Fiz. Pis'ma Red. 4 (1966) 144 .

\bibitem{Hillas84}
A. M. Hillas, Annu. Rev. Astron. Astrophys. 22 (1984) 425.

\bibitem{Berezinsky97}
V. S. Berezinsky, M. Kachelriess, A. Vilenkin, Phys. Rev. Lett. 79
(1997) 237.

\bibitem{auger}
Auger Collaboration, Nucl. Inst. and Meth. A 523 (2004) 50 .

\bibitem{EUSO}
EUSO Collaboration, www.euso-mission.org.

\bibitem{Billoir2000}
P. Billoir, Auger technical note GAP-00-025 (2000).

\bibitem{Bird95}
J. D. Bird et al., Astrophys. J. 441 (1995) 144.

\bibitem{Halzen95}
F. Halzen, A. R. Vazquez, T. Stanev, H. P. Vankov, Astropar. Phys.
3 (1995) 151.

\bibitem{Nagano99}
M. Nagano et al., Astropart. Phys. 13 (2000) 4.

\bibitem{Ave00}
M. Ave et al., Phys. Rev. Lett. 85 (2000) 2244.

\bibitem{aires0}
S. J. Sciutto, \emph{in Proceedings of 27th International Cosmic
Ray Conference}, Hamburg, Germany, 2001 (Copernicus Gesellshaft,
Katlemburg-Lindau, 2001) p 237; avaliable electronically,
www.fisica.unlp.edu.ar/auger/aires

\bibitem{QGSJET}
N. N. Kalmykov, S. S. Ostapchenko, A. I. Pavlov, Nucl. Phys.
(Proc. Suppl.) 52B (1997) 17.

\bibitem{SIBYLL}
R. S. Fletcher, T. K. Gaisser, P. Lipari, T. Stanev, Phys. Rev. D
50 (1994) 5710.

\bibitem{Hassoun95}
M. H. Hassoun, Fundamentals of Artificial Neural Networks (1995)
(The MIT Press, Massachusetts, U.S.).

\bibitem{GMTSci2001}
G. A. Medina-Tanco, S. J. Sciutto, \emph{in Proceedings of 27th
ICRC}, Hamburg, Germany, 2001 (Copernicus Gesellshaft,
Katlemburg-Lindau, 2001) p 518.

\bibitem{rumc86}
D. E. Rummelhart, J. McCelelland, Parallel Distributed Processing
(1986) (The MIT Press, Cambridge, Massachusetts).

\bibitem{Demuth99}
H. Demuth, M. Beale, Neural Network Tool Box -MATLab - Use's Guide
(1999), verson 4.

\bibitem{foha97}
F. D. Foresee, and M. T. Hagan, \emph{ in Proceedings of the 1997
Inter. Joint Conf. on Neural Network} (1997) pp 1930-1935.

\bibitem{hade96}
M. T. Hagan, H. B. Demuth, M. Beale, Neural Network Design (1996),
(PWS Publishing Company, Boston,).

\bibitem{GMTSciTiba2003}
G. A. Medina-Tanco, S. J. Sciutto, A. K. O. Tiba, \emph{in
Proceedings of 28th International Cosmic Ray Conference}, Tsukuba,
Japan, (2003).

\bibitem{Almu04}
J. Alvarez-Mu\~{n}iz et al., Phys. Rev. D 69 (2004) 103003.

\end{thebibliography}

\end{document}